%% file: main.tex
\documentclass[conference]{IEEEtran}

\usepackage{cite}
\usepackage{setspace}
\usepackage{graphicx}
\usepackage{array}
\usepackage{tikz}
\usepackage{tikz-3dplot}
\usetikzlibrary{shapes.geometric}
\usetikzlibrary{decorations.pathreplacing}
\usetikzlibrary{positioning}
\usetikzlibrary{patterns}
\usepackage{pgfplots}
\usepgfplotslibrary{fillbetween}
\usepackage{pstool}  
\usepackage{mathtools}
\usepackage[top=0.7in, bottom=1.03in, left=0.7in, right=0.7in]{geometry}  
\usepackage[linesnumbered,algoruled,boxed,lined]{algorithm2e}
\usepackage[absolute]{textpos}
\usepackage{threeparttable}
\usepackage[strict]{changepage}
\usepackage{flushend} 
\usepackage{amssymb}
\usepackage{comment}
\usepackage{bbm}
\usepackage{hhline}
\usepackage{lipsum} 
\usepackage{enumitem}
\usepackage{soul}
\usepackage{xcolor}

\usepackage{amsthm}

\ifCLASSINFOpdf
\else
\fi
\usepackage{amsmath}
\ifCLASSOPTIONcompsoc
  \usepackage[caption=false,font=normalsize,labelfont=sf,textfont=sf]{subfig}
\else
  \usepackage[caption=false,font=footnotesize]{subfig}
\fi
\pgfplotsset{compat=1.14} 

\begin{document}
%
\title{Uncoordinated Interference Avoidance Between Terrestrial and Non-Terrestrial Communications}
%


\author{\IEEEauthorblockN{Faris B. Mismar\IEEEauthorrefmark{1} and Aliye \"{O}zge Kaya\IEEEauthorrefmark{2}}
\IEEEauthorblockA{\IEEEauthorrefmark{1} Nokia Bell Labs Consulting, Murray Hill, NJ, 07974}%
\IEEEauthorblockA{\IEEEauthorrefmark{2} Nokia Bell Labs, Murray Hill, NJ, 07974}
}

\maketitle
\begin{abstract}
This paper proposes an algorithm that uses geospatial analytics and the muting of physical resources in next-generation base stations (BSs) to avoid interference between cellular (or terrestrial) and satellite communication (non-terrestrial) systems.   The information exchange between satellite and terrestrial stations is minimal, but a hybrid edge cloud node with access to estimated satellite trajectories can enable these BSs to take proactive steps to avoid interference.  To validate the superiority of our proposed algorithm over a conventional method, we show the performance of the algorithm using two measures: number of concurrent uses of Doppler corrected radio frequency resources and the sum-rate capacity of the BSs.  Our algorithm not only provides significant sum-rate capacity gains in both directions enabling better use of the spectrum, but also runs in polynomial time, making it suitable for real-time interference avoidance.
\end{abstract}



%
\IEEEpeerreviewmaketitle

\section{Introduction}\label{sec:introduction} 
%
%
%
%



 
Since the introduction of the fifth generation of wireless communications (5G), the competition towards wireless spectrum has become ever challenging as cellular operators desire to harness more spectrum.  This is necessary to meet the increasing demand for high speed data.  Given that spectrum is a scarce resource that is pre-allocated to incumbent industries besides cellular, including non-terrestrial communication systems, interference becomes inevitable.


Satellites and in particular low Earth orbit (LEO) satellites possess wide-area coverage with significantly
reduced latency due to lower altitude orbits (e.g., compared to geosynchronous satellites).  However, due to that low orbit, the velocity of these non-terrestrial objects is very high and thus each LEO satellite can make several Earth turns per day.  Thus, even if a terrestrial base station (BS) is granted access to transmit at a frequency that belongs to the satellite frequency range, it has a short period of time to transmit, or else it will interfere with the satellite transmission.  Therefore, avoiding the interference allows the BS to transmit for longer periods through improved schemes of spectrum sharing.  


There is no shortage of research about interference \textit{detection} be it between satellites and terrestrial nodes or within terrestrial nodes such as BSs \cite{7833539, 7928993, 9693912, 9860900}.  For example, an interference detection scheme was proposed between satellites and terrestrial links in \cite{7833539}, where this was done through measuring saturation at the receiver chain.  Then, digital filtering techniques were applied to mitigate these interfering signals.  The methodology we propose is for interference \textit{avoidance}, where we apply geospatial techniques and frequency corrections due to the difference in velocity between LEO satellites and terrestrial BSs, and then block the radio resources that could cause the interference.  These resources are later unblocked when interference is no longer a problem.  In \cite{7928993}, it was asserted that the information exchange between satellite and terrestrial links was very limited.  Therefore, obtaining a full channel state information (CSI) for interference mitigation was almost impossible.  We propose a solution to the  problem by avoiding the interference without having to exploit the CSI.%

\begin{figure}[!t]
\centering
\resizebox{0.4\textwidth}{!}{
    \psfragfig{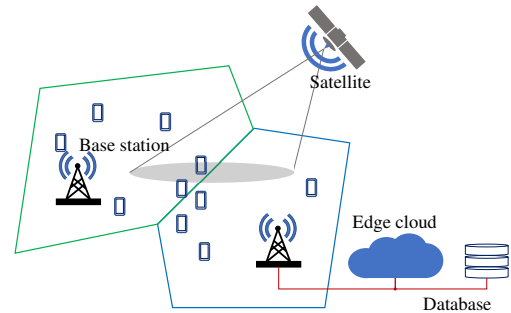}{%
      \psfrag{B}[][][2.2]{\Large Base station}%
      \psfrag{C}[][][2.2]{\Large Edge cloud}%
      \psfrag{S}[][][2.2]{\Large Satellite}%
      \psfrag{D}[][][2.2]{\Large Database}%
    }
}
\caption{A satellite covering a next-generation base station association area.}
\label{fig:overall}
\end{figure}%

The use of edge nodes or hybrid cloud edge nodes in next-generation radio access networks (RAN) continues to proliferate \cite{9500204, arxiv220911312}.   In the area of real-time performance self-diagnosis of faults and anomalies, an algorithm was proposed in \cite{9500204}.  This algorithm uses either an edge node, which we consider, or the real-time Open-RAN intelligent controller (RIC).  The use of a hybrid cloud edge node helps collect data about satellite movements, often tracked in a government managed database, in addition to the ability to perform computations as required by an algorithm, such as the one we propose.  An architecture that exploits this setup is shown in Fig.~\ref{fig:overall}.


From an industry standards point \cite{3gpp38863}, a study was made on the co-existence of both terrestrial and non-terrestrial (i.e., satellite) communications.  Several scenarios were considered with the focus on the throughput loss as a result of the adjacent channel interference ratio as the performance measure.  However, unlike our paper, the study did not propose any methods to mitigate or avoid interference on the satellite links.


In this paper, we propose a solution to the interference that the BSs cause onto the satellite-Earth transmission in shared frequency bands.  The main contributions of this paper can be summarized as follows:

\begin{enumerate}
\item Formulate the problem of interference avoidance when transmission is uncoordinated between satellites and base stations.
\item Exploiting muting 5G physical resource blocks (PRB), known as ``PRB blanking'', in addition to geospatial techniques to mitigate the resulting interference.
\end{enumerate}%

\vspace*{-0.5em}
\section{System Model and Problem Formulation}\label{sec:sysmodel}
This section presents a system model for coexisting cellular and satellite communications.  We formulate the problem of interference avoidance due to terrestrial and satellite simultaneous communications as an optimization problem.

\subsection{System Model}
 
 The system comprises two communication subsystems: 1) a sectorized terrestrial cellular system that uses beamforming with $M \ge 1$ transmit antennas per sector and PRB resources and 2) a satellite communication system communicating towards the Earth on a carrier the frequency of which coincides with a PRB.  A PRB is the smallest non-overlapping element of resource allocation assigned by the BS scheduler over a finite time slot duration.  Let $\mathcal{B}$ be the set of BSs and $\mathcal{L}$ be the set of satellites in the system.  Thus, we write the \textit{downlink} system model as follows:
\begin{equation}\label{eq:system}
y_{b,k} = \sqrt{G_b} \mathbf{h}_{b,k}^\ast \mathbf{f}^\star_{b,k} 
x_{b,k}  + \underbrace{\sum_{b^\prime \in \mathcal{E}} \sqrt{G_{b^\prime}}\mathbf{h}_{{b^\prime},k}^\ast 
\mathbf{f}^\star_{b,k} x_{{b^\prime},k}}_{\coloneqq i_k} + n_{b,k}
\end{equation}
where $y_{b,k}$ ($x_{b,k}$) is the received (transmitted) signal on the $k$-th PRB $k\in\mathcal{K} \coloneqq \{1,2,\ldots,N_\text{PRB}\}$ from the serving BS and interfering satellite, $b$ and $s$, respectively.  $\mathbf{h}_{\cdot,k}$ is power-normalized channel with a large-scale fading coefficient $G_\cdot$, and $\mathbf{f}_{b,k}^\star$ is the optimal beamforming vector, which is
\begin{equation}
\mathbf{f}_{b,k}^\star \coloneqq \underset{\mathbf{f}\in\mathcal{F}}{\arg\,\max}\; \vert \sqrt{G_b} \mathbf{h}_{b,k}^\ast  \mathbf{f} \vert^2
\end{equation}%
where $\mathcal{F}$ is a pre-defined codebook of beams (i.e., a grid of beams for a total of $\vert \mathcal{F} \vert$ beams), such that each beamforming vector $\Vert \mathbf{f} \Vert^2 = M$, $x_{\cdot,k}$ is the transmitted signal from the BS $b$ on the $k$-th PRB or the interfering satellite $s$ transmission on a frequency coinciding with the $k$-th PRB, $i_k$ is the interference term on the $k$-th PRB, and $n_k\sim\text{Norm}(0,\sigma^2)$ independently sampled from a complex Normal distribution.  The transmit power per PRB is $0 \le P^{(b, k)}_x \le P_\text{max} \coloneqq P_\text{BS}/N_\text{PRB}$ for the downlink with $P_\text{UE}$ replacing the BS transmit power for the uplink. This model holds true regardless of the number of user equipment (UEs) due to the scheduling operation from the serving BS.  Further, let us define the set of all interferers $\mathcal{E}\coloneqq \mathcal{L}\cup \mathcal{B}\setminus \{b\}$ which includes the interfering satellite as well as any terrestrial BS in the system excluding the serving BS. Next, the received signal to interference plus noise (SINR) of a given PRB $k$ sent by a base station $b$ is 
\begin{equation}\label{eq:sinr}
\textsf{SINR}_{b, k} = \frac{G_b P_{x}^{(b, k)} \mathbb{E} \vert \mathbf{h}_{b,k}^\ast  \mathbf{f}_{b,k}^\star \vert^2}{\sigma^2 + \sum_{{b^\prime}\in \mathcal{E}} \mathbb{E} \vert \sqrt{G_{b^\prime}} \mathbf{h}_{{b^\prime},k}^\ast\mathbf{f}_{b,k}^\star x_{{b^\prime},k}  \vert ^2}
\end{equation}%

The system model on the uplink is analogous to \eqref{eq:system} where the BS $b$ is replaced with a UE $u$.  The satellite; however, continues to transmit on the downlink.  In the uplink, the BS controls which UE transmits on the $k$-th PRB to the serving BS by means of a scheduling request on the physical uplink control channel.  The channel reciprocity is irrelevant to the problem statement since our focus is on the interference as a statistic.  Nevertheless, in time division duplex (TDD) systems, where exploiting channel reciprocity is possible, the uplink channels are equal to $\mathbf{h}_{\cdot,k}$.  For frequency division duplex (FDD) systems, different channel and beamforming vector notations can be used for the uplink system model, but the underlying principles are similar.

\subsection{Problem Statement}
The problem of avoiding the uncoordinated interference from the satellite subsystem can be written as an optimization problem, with the total interference power being the optimization objective:
\begin{equation}\label{eq:problem}
\begin{aligned}
\text{minimize:} \qquad & \sum_{k} \sum_{b^\prime\in\mathcal{E}} \mathbb{E} \vert \sqrt{G_b^\prime} \mathbf{h}_{b^\prime,k}^\ast\mathbf{f}_{b,k}^\star {x}_{b^\prime,k} \vert ^2 \\
\text{subject to:} \qquad & 0 \le \mathbb{E} \vert {x}_{b^\prime,k} \vert ^ 2 \le P_\text{max}, \\ 
&  \mathbf{f}^\star_{b,k} \in \mathcal{F}, \\
& k \in \{1, 2, \ldots, \max(\mathcal{\tilde K})\} \\
\end{aligned}
\end{equation}
where $\mathcal{\tilde K} \subseteq \mathcal{K}$ is the subset of PRBs that are utilized at a given traffic load served by the BS.  The first condition constrains the power per resource (i.e., through power control power increments or decrements).   This problem is not a convex problem since the optimization set is not a closed set (due to the discrete sets in the constraints).  The utilization for a given BS is the ratio $\vert\mathcal{\tilde K}\vert$ / $\vert\mathcal{K}\vert$.

\vspace*{-0.25em}
\section{Uncoordinated Interference Avoidance}
This section describes the different considerations for coverage from both radio frequency and geospatial points of view.  We then delve into explaining our proposed algorithm to avoid uncoordinated interference.
\subsection{Frequency Allocation}
Satellite carriers over which the signal is transmitted can be mapped to specific PRBs.  The conversion from a carrier frequency to a PRB given the operating band and direction are known is straightforward (and so is the conversion from a PRB to a carrier frequency) \cite{3gpp38104}.  The equation that converts the $k$-th PRB to its beginning frequency $f(k; \cdot)$ is given by:
\begin{equation}
f(k; B, N_\text{PRB}) = f_\text{start}^B +\frac{f_\text{end}^B - f_\text{start}^B}{N_\text{PRB}} \cdot (k - 1),
\label{eq:prb}
\end{equation}
where $f_\text{start}$ and $f_\text{end}$ denote the start and end \textit{usable} frequencies (i.e., excluding the guard bands) 
and the PRB range is $k \in \{1,2,\ldots, N_\text{PRB}\}$.  The equation applies to both uplink and downlink direction.  However, a duplex distance $f_\text{duplex}^\text{n70}$ has to be taken into consideration.  This duplex distance is the frequency separation between the transmit frequency and the receive frequency.  Since PRBs have a bandwidth of subcarrier spacing $\Delta f$ multiplied by the number of subcarriers per PRB $N_\text{SC}$, the ending frequency of a given PRB can be found by adding the term $N_\text{SC} \cdot \Delta f$ to the frequency computed in \eqref{eq:prb}.  From the perspective of a satellite equipment, the frequency has to be corrected due to the BS-relative rotational speed of LEO satellites, which causes a Doppler effect.

\textbf{Frequency correction:} For satellite communications with stationary objects, Doppler correction is necessary. A finite quantity $f_\text{correction}$ is added to (or subtracted from) the \textit{transmitted} frequency based on whether the satellite projection is approaching (or departing) the BS:
\begin{equation}\label{eq:doppler}
   f_\text{correction} = \frac{v_s}{c} f_0 
\end{equation}
where $v_s$ is the satellite speed derived from the altitude, $c$ is the speed of light, and $f_0$ is the transmit sky frequency.  The corrected frequency at the receiver is thus $f = f_0 \pm f_\text{correction}$.  If we define the \textit{estimated} Euclidean distance between the satellite projection and the sector $D(t + 1; q,v)\coloneqq \Vert \mathbf{p}^\prime_v - \mathbf{p}_q \Vert$, where $\mathbf{p}_\cdot$ is the position vector (as defined later) at time $t + 1$, then based on the rate of change in the distance $\Delta D$/$\Delta t$, we state if negative (positive), then the satellite is approaching (departing) the BS.
The choice of FDD or TDD impacts the start and end frequencies in \eqref{eq:prb}.  For example, in the NR band n70, which is a FDD band in the sub-6 GHz range, the duplex distance is $f_\text{duplex}^\text{n70} = 300$ MHz \cite{3gpp38104}.  This means that if the downlink starting frequency was $f_0$, then the \textit{uplink} starting frequency would be $f_0 - 300$ MHz.  Comparably, NR band n77 is a TDD band in the sub-6 GHz range, which coincides with the satellite C~band.  The uplink and downlink frequencies are similar but are separated in time based on the NR radio frame structure and configuration \cite{3gpp38211}.%
\vspace*{-0.25em}
\subsection{Cellular Coverage}
For the geometry of cellular coverage, we use the Voronoi tessellation  algorithm.  Every BS in $\mathcal{B}$ has three sectors the set of which is $\mathcal{Q}$.  Each sector $b$ constructs its own longitude-latitude point $\mathbf{p}_b$ for a total of $\vert\mathcal{Q}\vert = 3\vert\mathcal{B}\vert$ points scattered in a two-dimensional geographical area.   The Voronoi tessellation algorithm partitions this two-dimensional plane into Voronoi \textit{cells}, each of which is $\mathcal{R}_q$, where $q > 0$ is the number of cells.  Using Voronoi tessellation is an acceptable geospatial practice in finding the \textit{natural} service area of a BS \cite{Okabe2000SpatialTC, 6576422}.  Since each sector is assigned to one Voronoi cell, $q = \vert \mathcal{Q}\vert= 3\vert \mathcal{B}\vert$. 


\tikzset{reverseclip/.style={insert path={(current bounding box.south west)rectangle (current bounding box.north east)} }}
\begin{figure}[!t]
\centering
\resizebox{0.35\textwidth}{!}{
    \psfragfig{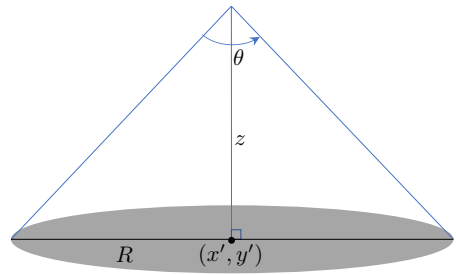}{%
      \psfrag{t}[][][1.4]{$\theta$}%
      \psfrag{o}[][][1.4]{$(x^\prime, y^\prime)$}%
      \psfrag{z}[][][1.4]{$\Large z$}%
      \psfrag{R}[][][1.4]{$\Large R$}%
    }
}
\caption{Projection of a satellite coverage onto the Earth.}
\label{fig:projection}
\end{figure}%

Owed to the Voronoi tessellation algorithm, each cell consists of every point in the two-dimensional plane whose Euclidean distance to a given BS sector is less than or equal to its distance to any other BS sector.  Formally
\begin{equation}\label{eq:space}
    \mathcal{R}_b \coloneqq \{\mathbf{r} \in\mathcal{A} \mid \Vert \mathbf{r} - \mathbf{p}_b \Vert \le \Vert \mathbf{r} - \mathbf{p}_{b^\prime}\Vert, \forall b\neq b^\prime) \}
\end{equation}
where $\mathcal{A}$ is a metric space corresponding to the two-dimensional geographical area. 


\subsection{Satellite Trajectory}
Satellites, and particularly low Earth orbit (LEO) satellites, may use certain frequencies for satellite communications that interfere with cellular communication.  However, since the time-based trajectories of satellites are known to a degree of certainty, computing their coverage projection on to Earth is possible.  This computation, when carried out for a given altitude, allows us to also compute whether or not interference is likely (i.e., due to overlapping coverage areas of the satellite projections onto the Voronoi cells of the BSs). 

Let there be a non-empty set of satellites $\mathcal{E}$.  The trajectory of a given satellite at a given time instance $t$ is given by the column vector $\mathbf{p}^\top(t)\coloneqq[x(t), y(t), z(t)]$, which denotes the longitude, latitude, and altitude, all at the time $t$ respectively. A simplified approach of 1) using circular geometry of the satellite coverage area and 2) assuming a flat surface on Earth \cite{6558007} allows us to compute the projection of the satellite coverage on to Earth.  In effect, if the LEO is at an altitude $z(t) = \ell$ from the surface of Earth, and the satellite equipment has an antenna with a beamwidth of $\theta$ as shown in Fig.~\ref{fig:projection}, then the radius of the projected circular geometry of the satellite coverage can be computed as%
\begin{equation}
    \label{eq:projection}
    R = \frac{1 - \cos\theta}{\sin\theta}\cdot \ell 
\end{equation}
which enables us to identify the projection region of the $v$-th satellite $\mathcal{S}_v, v\in\mathcal{E}$ at a given time through a projected center $(x^\prime,y^\prime)$ and a radius $R$:
\begin{equation}
    \mathcal{S}_v\coloneqq\{\mathbf{p}_v^\prime \coloneqq (x^\prime,y^\prime)\mid \Vert \mathbf{r} -\mathbf{p}_v^\prime \Vert^2 \le R^2, \forall \mathbf{r} \in \mathcal{A} \}
\end{equation}
where the time instance $t$ is dropped for clarity and $\mathcal{A}$ is the same metric space as defined in \eqref{eq:space}. %

\subsection{Algorithm Outline}\label{sec:algorithm_outline}
At this stage what is left is to compute whether the projection of any given satellite would overlap with the coverage area of any BS sector.   If this holds true, then interference is possible.  To avoid it, the said BS \textit{proactively} ceases the transmission (or reception) on the $k$-th downlink (or uplink) PRB of the given sector.  This is done for any sector-satellite pair in the area.  To cease transmission or reception on a given PRB, the PRB blanking algorithm \cite{patent} is invoked where a PRB is technically disabled from transmission or reception.  Reversing the blanking (i.e., unblanking) can take place only if there is no computed possibility of interference.  That is, allow a PRB to be reused for transmission or reception only if no satellite overlap with the serving BS sector is computed.  
This is necessary to prevent interference towards the satellite links, which could have public safety violations.  
The details are outlined in Algorithm~\ref{alg:algorithm}.

\SetKwFor{Loop}{Loop}{}{EndLoop}
\begin{algorithm}[!t]
    \small
    \caption{Uncoordinated Interference Avoidance}
    \label{alg:algorithm}
    \DontPrintSemicolon
    \KwIn{1) Coordinates of BSs including sectorization, frequencies, and coordinates and 2) frequencies of LEO satellites as a function of time.}
    \KwOut{Blocking and unblocking sequence of resources per sector for all BSs over time.}
    Pre-compute the Voronoi tessellation for all BSs $\mathcal{R}_q$.\;
    $t \gets 0$ \;
    \Loop{} {
        Read the trajectory database for satellite $v$. \;
        \ForEach {$v \in \mathcal{E}$} {
            Compute the projection region $\mathcal{S}_v$.\;
            \ForEach {$q \in \mathcal{Q}$} {
                $o_q \gets O(\mathcal{R}_q, \mathcal{S}_v)$ \;
                \eIf {$o_q = 1$} { 
                    Compute $D(t + 1;q,v)$ and find whether satellite $v$ will be approaching or departing $q$. \;
                    Correct the frequencies used by sector $q$ (both directions) due to Doppler as perceived by the satellite $v$ using \eqref{eq:doppler}.\;
                    Convert corrected frequencies to PRB(s) using \eqref{eq:prb}. \;
                    \If {$\text{\rm PRB(s) blankable}$} {
                        Proactively blank PRBs corresponding to frequency and direction. \;
                    }
                } 
                {
                    Unblock PRBs on $q$. \;
                }
            } 
        } 
        $t \gets t + 1$ \;
    }
\end{algorithm}%
\textbf{Overlap computation:} To compute whether an overlap is possible, the intersection of the metric spaces corresponding to the $q$-th BS sector and the $v$-th satellite equipment: %
\begin{equation}\label{eq:overlap}
    O(\mathcal{R}_q, \mathcal{S}_v) \coloneqq \mathbbm{1}[\mathcal{R}_q \cap \mathcal{S}_v \neq \varnothing] 
\end{equation}
which is equal to zero when there is no overlap between the coverage areas of the two.

\textbf{Run-time complexity:} Since the run-time complexity of Voronoi tessellation for $q$ cells is in $\mathcal{O}(q\log q)$\cite{aggarwal}, the run-time complexity of the algorithm is in $O(q\log q + q\vert \mathcal{L}\vert)$, which is polynomial time in $q$.   This holds true as long as the sectorization of the BSs continuously change (e.g., by means of automated antenna azimuth optimization).  However, if sectorization is fixed, then the Voronoi computation step in the algorithm is run only once bringing the run-time complexity to $\mathcal{O}(q\vert\mathcal{L}\vert)$ in subsequent runs.
\section{Simulation}\label{sec:simulation}%
This section describes the simulation setup, proposed performance measures, and a discussion about results.%
\subsection{Setup}

We consider a \textit{realistic}\footnote[3]{True coordinates have been translated to conceal the original locations.}
geographical rural area covered by five macro BS sites.  Each BS has three sectors with an inter-site distance range of $3.90$ to $5.21$ km.  Each sector has a bandwidth of $N_\text{SC}\cdot\Delta f = 180$ kHz per PRB.  The parameters are outlined in Table~\ref{tab:simulation}.  Three LEO satellites hover over this area during a specific time duration.  PRBs are scheduled on the uplink and downlink independently.   We consider a sufficient number of UEs that scattered uniformly across the service areas of each sector.  Further, we follow a grid of beams approach for the beamforming codebook $\mathcal{F}$ with $M = 4$ transmit antennas. To ensure adequate utilization, we consider a full-buffer traffic model in both directions.%
\subsection{Performance Measure}
\textbf{Collisions:} To measure the performance of the algorithm, we define a  ``collision'' as a moment where 1) an overlap of a satellite projection onto the Voronoi cell (or union of cells thereof) of a BS sector (or several sectors thereof), defined in \eqref{eq:overlap} and 2) more than one transmission takes place simultaneously within the frequency band of one PRB (i.e., due to a transmission from a BS (or UE) that coincides with the corrected sky frequency (i.e., from a satellite).  Clearly, a collision is an indication of uncoordinated interference.  If we denote the Doppler corrected frequency transmitted by the satellite $v$ as $f_\text{corrected}^{(v)}$, as computed in \eqref{eq:doppler}, then the number of collisions across the network is given by
\begin{equation}\label{eq:collision}
\begin{aligned}
& C(\mathcal{B}, \mathcal{L}) = \\
& \sum_{q\in\mathcal{B}, v\in\mathcal{L}} \sum_k O(\mathcal{R}_q, \mathcal{S}_v)\cdot  \mathbbm{1}[f(k^{(q)}; B^{(q)}, N_\text{PRB}^{(q)}) = f_\text{corrected}^{(v)}].
\end{aligned}
\end{equation}

Intuitively, minimizing the objective in \eqref{eq:problem} implies minimizing collisions per direction (i.e., uplink or downlink) since they are the only source of interference in the problem statement.  This is because no BS or UE pairs can transmit to the same PRB at the same time and space, as this is handled within NR itself through control channel signaling \cite{3gpp38300}.

\textbf{Sum-rate capacity:} To further measure the performance of the terrestrial network in the presence of PRB blocking, we introduce the sum-rate capacity defined for the sectors that witnessed a collision around the time of that collision  $\mathcal{Q}^{(C > 0)}$:
\begin{equation}\label{eq:sumrate}
\begin{aligned}
    S(\mathcal{Q}^{(C > 0)}, \mathcal{\tilde K}) & \coloneqq \sum_{q\in\mathcal{Q}^{(C > 0)}} \sum_{k \in \mathcal{\tilde K}(q)} W(k;q) \\
    &\leq \sum_{q\in\mathcal{Q}^{(C > 0)}} \sum_{k \in \mathcal{\tilde K}(q)} \log_2(1 + \textsf{SINR}_{b, k} ) \\
\end{aligned}
\end{equation}%
where $W(k;q)$ is the information rate as measured for the $k$-th PRB in the $q$-th sector in bits per channel use (c.u.) units.  

To establish a baseline for benchmarking purposes, we run a simulation with our proposed algorithm turned off.  This is equivalent to the practically widely adopted equal power allocation (EPA) scheme.  EPA in essence would mean that the BS power would be uniformly allocated across all PRBs (i.e., no blanking of PRBs due to satellites overlapping can occur).  Thus at full load, interference from a terrestrial BS on satellite transmission is almost certain due to simultaneous allocation of PRBs that coincide with a satellite transmission.

\begin{table}[!t]
\centering
\setlength\doublerulesep{0.5pt}
\caption{Simulation settings}
\vspace*{-0.5em}
\label{tab:simulation}
\small
\begin{tabular}{ ll } 
\hhline{==}
Frequency band & C band  \\
Satellite beamwidth & $60^\circ$ \\
Number of PRBs $N_\text{PRB}$ & 50 \\
BS sector utilization & $\{10, 30, \ldots, 100 \}\%$ \\
BS maximum power $P_\text{BS}$ & 40 W \\
UE maximum power $P_\text{UE}$ & 200 mW \\
LEO satellite Earth speed & $7{,}800$ m/s \\
\hhline{==}
\end{tabular}
\end{table}%

\subsection{Results}
\begin{figure}[!t]
\centering
\resizebox{0.47\textwidth}{!}{\includegraphics[]{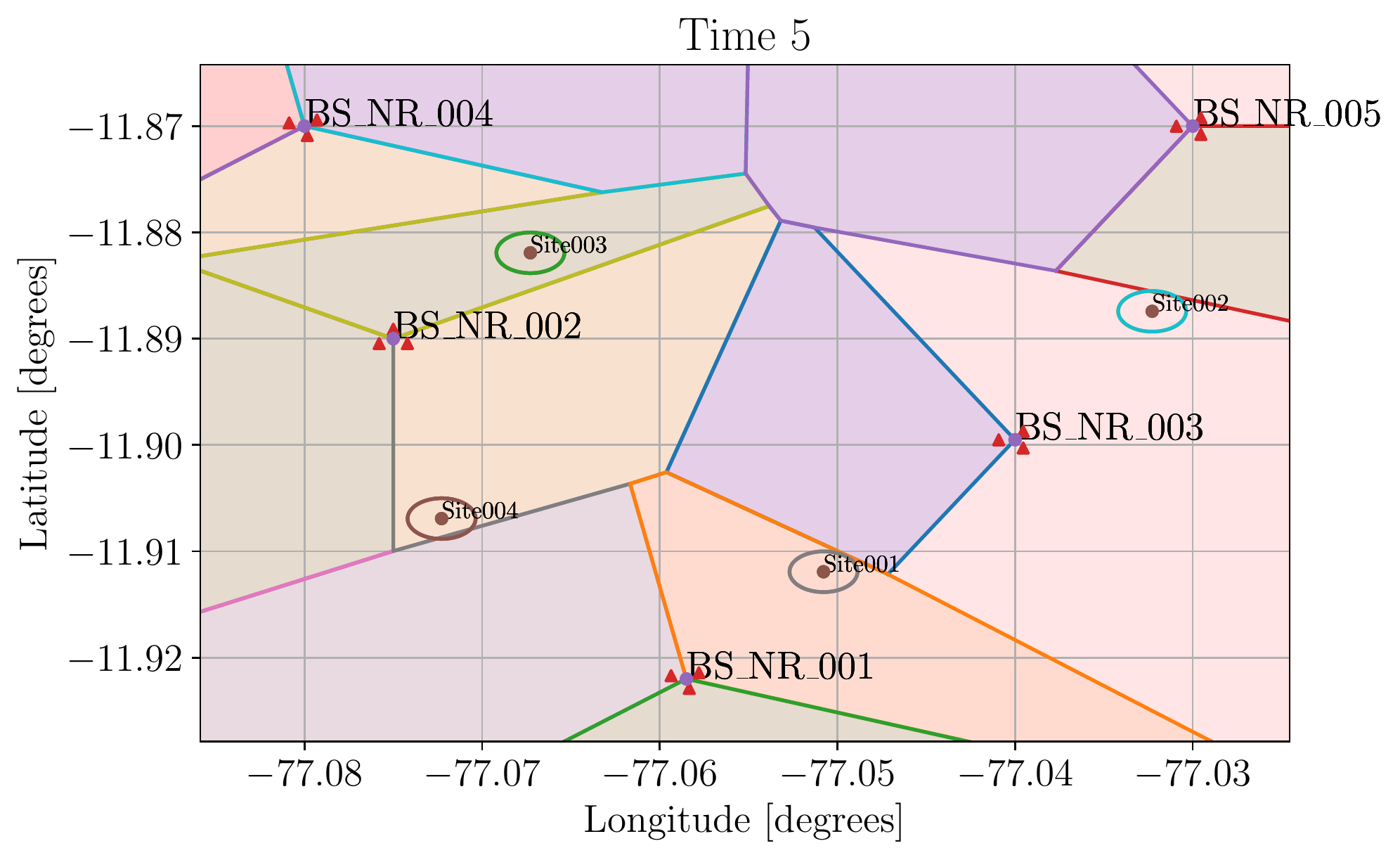}}
\caption{Projection of satellites coverage onto an area served by sectorized~BSs.}
\label{fig:simulation}
\end{figure}

\begin{figure}[!t]
\centering
\resizebox{0.4\textwidth}{!}{\input{figures/cdf_ul.tikz}}
\caption{Empirical cumulative distributions of the uplink sum-rate capacity for the two algorithms: proposed and EPA.}
\label{fig:result_3}
\end{figure}
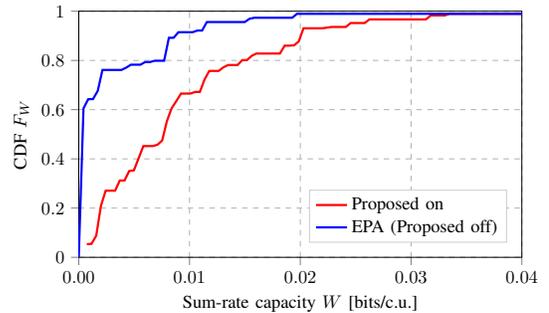

Fig.~\ref{fig:simulation} shows the simulation of our geographical area at a particular time instance.  Each sector of the three-sectored BSs generates a Voronoi cell, which is the natural service area of the sector, where UEs are uniformly scattered and LEO satellites can fly over.  Results are shown in Fig.~\ref{fig:result_1} to Fig.~\ref{fig:result_3}.  

\begin{figure}[!t]
\centering
\resizebox{0.4\textwidth}{!}{\input{figures/results.tikz}}
\caption{Number of collisions for the two algorithms: proposed and EPA.}
\label{fig:result_1}
\end{figure}
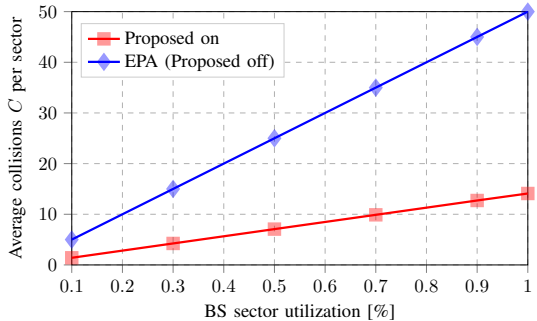%

\begin{figure}[!t]
\centering
\resizebox{0.4\textwidth}{!}{\input{figures/cdf_dl.tikz}}
\caption{Empirical cumulative distributions of the downlink sum-rate capacity for the two algorithms: proposed and EPA.}
\label{fig:result_2}
\end{figure}
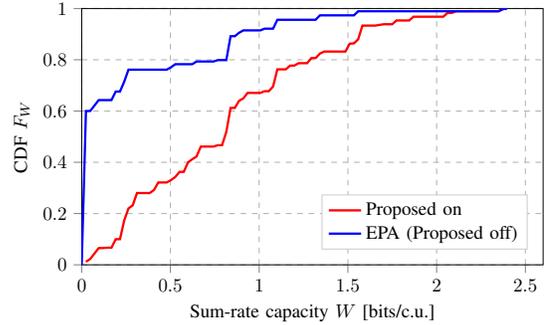%

\subsection{Discussion}
We discuss two questions stemming from observing the plot in Fig.~\ref{fig:result_1} and Fig.~\ref{fig:result_2}: 1) Why is the average number of collisions non-zero for our proposed algorithm? 2) What is the effect of our proposed algorithm on the performance of the terrestrial subsystem as measured by the sum-rate capacity in both the downlink and the uplink?

To answer the first question, we consider the scenario where a group of UEs are served by a BS sector the realistic coverage boundaries of which do not coincide with the Voronoi cells boundaries.  This can happen due to non-uniformity in propagation losses that impact the selection of the serving sector based on non-distance measures (i.e., signal strength or quality).  Thus, the blanked PRB may not have been the optimal one to blank, causing a collision.  
A secondary question from this main question arises: Why does the performance disparity, measured by the gap between the two algorithms, increase non-uniformly as the utilization increases?  We keep in mind that at full buffer, all allocated PRBs are used in transferring data. For EPA, once a satellite passes, all allocated PRBs cause collisions.  However, for our proposed algorithm,  PRBs are proactively blanked minimizing the instantaneous collisions, which reduces the average overall.%

In answering the second question, we study the cumulative distribution function (CDF) $F_W(\cdot)$ of the sum-rate capacity \eqref{eq:sumrate}.  We do this in each transmission direction (i.e., uplink and downlink) and for both algorithms around times of collision.  Fig.~\ref{fig:result_2} and Fig.~\ref{fig:result_3} show these CDFs. We start with the observation that in the EPA algorithm, the sum-rate capacity of the system has a high probability mass at $W = 0$ for both transmission directions.  This means that the transmitting BS is in outage due to collisions with satellite transmissions that represent approximately $60\%$ of the samples collected during the simulation.  In our proposed algorithm, and due to the proactive blocking of the PRBs involved in the collision, a more effective use of the spectrum is demonstrated where $60\%$ of the samples have a downlink sum-rate capacity of $0.4$ bits/c.u.  Mathematically, $F_{W_\text{EPA}}^{-1}(0.6) \approx 0$ and $F_{W_\text{Proposed}}^{-1}(0.6) \approx 0.4$ in the downlink.  For the uplink, $F_{W_\text{EPA}}^{-1}(0.6) \approx 0$ and $F_{W_\text{Proposed}}^{-1}(0.6) \approx 0.008$.  We also observe that this performance improvement persists for any percentage of samples.  In fact, $F_{W_\text{EPA}}^{-1}(0.8) \approx 0.81$ and $F_{W_\text{Proposed}}^{-1}(0.8) \approx 1.29$ in the downlink an  $F_{W_\text{EPA}}^{-1}(0.8) \approx 0.007$ and $F_{W_\text{Proposed}}^{-1}(0.8) \approx 0.0147$ in the uplink.  Thus, a sum-rate capacity gain of $1.6$x ($2.1$x) in the downlink (uplink) is observed in $80\%$ of the samples owed to our proposed algorithm.  An increase in the sum-rate capacity translates into a more effective use of the spectrum that is available to the terrestrial BSs.

Another observation from the figures is that the sum rate capacity on the uplink is generally lower than that on the downlink regardless of the choice of the interference avoidance algorithm.  The reason for this is because in the uplink the UE is the transmitter, and its power is orders of magnitude lower than that of a BS.  In perspective, if a BS can transmit up to $40$ W at full load, a UE cannot transmit more than $200$ mW at full load to preserve its battery and adhere to body absorption rate limits.  Due to the constrained nature of the uplink transmission, a higher gain in the sum-rate capacity is even more beneficial to the system.%

\section{Conclusion}\label{sec:conclusion}
In this paper, we demonstrated the use of edge cloud enabled BSs and the exploitation of resource blocking as a viable means to handle uncoordinated interference between BSs and satellites as non-terrestrial systems.  We showed that colliding Doppler-corrected PRBs can be identified using geospatial analytics of the BS and satellite data.  The next-generation BSs could then cease transmission onto the resources minimizing the interference power towards these non-terrestrial systems.  This behavior can improve the sum-rate capacity of the terrestrial system in both uplink and downlink, leading to a more effective use of the spectrum overall. 

\bibliographystyle{IEEEtran}
\bibliography{main.bib}

\end{document}

%% file: figures/results.tikz
\begin{tikzpicture}

\begin{axis}[
width=4in,
height=2.5in,
legend cell align={left},
legend entries={Proposed on, EPA (Proposed off)},
legend style={at={(0.02,0.74)}, anchor=south west, draw=white!80.0!black, nodes={scale=1, transform shape}},
tick align=outside,
tick pos=left,
x grid style={white!69!black, dashed},
xlabel={BS sector utilization [\%]},
xmajorgrids,
xmin=0.1, xmax=1,
xtick={0,0.1,...,1},
y grid style={white!69!black, dashed},
ylabel={Average collisions $C$ per sector},
ymajorgrids,
ymin=0, ymax=50,
ytick={0,10,...,50.1},
]
\addplot [red, line width=1.2pt, mark=square*, mark size=3, mark options={solid, opacity=0.4}]
table [row sep=\\]{%
1   14.1\\
0.9 12.72\\
0.7 9.89\\
0.5 7.07 \\
0.3 4.24 \\
0.1 1.41 \\
};

\addplot [blue, line width=1.2pt, mark=diamond*, mark size=4, mark options={solid, opacity=0.4}]
table [row sep=\\]{%
1   50 \\
0.9 45\\
0.7 35\\
0.5 25\\
0.3 15\\
0.1 5\\
};
\end{axis}

\end{tikzpicture}

%% file: figures/cdf_dl.tikz
\begin{tikzpicture}

\begin{axis}[
width=4in,
height=2.5in,
legend cell align={left},
legend entries={Proposed on, EPA (Proposed off)},
legend style={at={(.52,0.06)}, anchor=south west, draw=white!80.0!black, nodes={scale=1, transform shape}},
tick align=outside,
tick pos=left,
x grid style={white!69!black, dashed},
xlabel={Sum-rate capacity $W$ [bits/c.u.]},
xmajorgrids,
xmin=0, xmax=2.6,
xtick={0,0.5,...,2.5},
y grid style={white!69!black, dashed},
ylabel={CDF $F_W$},
ymajorgrids,
ymin=0, ymax=1,
ytick={0,0.2, ..., 1.2}
]
\addplot [red, line width=1.2pt, 
mark options={solid, opacity=0.4}]
table {%
0.023967254370825 0.012
0.04793450874165 0.0226666666666667
0.071901763112475 0.044
0.0958690174832999 0.0653333333333334
0.119836271854125 0.0653333333333334
0.14380352622495 0.0666666666666667
0.167770780595775 0.0666666666666667
0.1917380349666 0.1
0.215705289337425 0.1
0.23967254370825 0.173333333333333
0.263639798079075 0.22
0.2876070524499 0.232
0.311574306820725 0.28
0.33554156119155 0.28
0.359508815562375 0.28
0.3834760699332 0.28
0.407443324304025 0.292
0.43141057867485 0.321333333333334
0.455377833045675 0.321333333333334
0.4793450874165 0.321333333333334
0.503312341787325 0.330666666666667
0.52727959615815 0.342666666666667
0.551246850528975 0.362666666666667
0.5752141048998 0.362666666666667
0.599181359270625 0.4
0.62314861364145 0.412
0.647115868012274 0.422666666666667
0.6710831223831 0.461333333333334
0.695050376753924 0.461333333333334
0.719017631124749 0.461333333333334
0.742984885495574 0.461333333333334
0.766952139866399 0.466666666666667
0.790919394237224 0.466666666666667
0.814886648608049 0.518666666666667
0.838853902978874 0.612
0.862821157349699 0.612
0.886788411720524 0.64
0.910755666091349 0.649333333333334
0.934722920462174 0.670666666666667
0.958690174832999 0.670666666666667
0.982657429203824 0.670666666666667
1.00662468357465 0.670666666666667
1.03059193794547 0.677333333333334
1.0545591923163 0.677333333333334
1.07852644668712 0.696
1.10249370105795 0.762666666666667
1.12646095542877 0.762666666666667
1.1504282097996 0.762666666666667
1.17439546417042 0.777333333333334
1.19836271854125 0.777333333333334
1.22232997291207 0.786666666666667
1.2462972272829 0.786666666666667
1.27026448165372 0.786666666666667
1.29423173602455 0.806666666666667
1.31819899039537 0.806666666666667
1.3421662447662 0.824000000000001
1.36613349913702 0.832000000000001
1.39010075350785 0.832000000000001
1.41406800787867 0.832000000000001
1.4380352622495 0.832000000000001
1.46200251662032 0.832000000000001
1.48596977099115 0.832000000000001
1.50993702536197 0.862666666666667
1.5339042797328 0.864
1.55787153410362 0.88
1.58183878847445 0.933333333333333
1.60580604284527 0.933333333333333
1.6297732972161 0.933333333333333
1.65374055158692 0.933333333333333
1.67770780595775 0.936
1.70167506032857 0.938666666666667
1.7256423146994 0.938666666666667
1.74960956907022 0.938666666666667
1.77357682344105 0.953333333333333
1.79754407781187 0.953333333333333
1.8215113321827 0.953333333333333
1.84547858655352 0.953333333333333
1.86944584092435 0.968
1.89341309529517 0.968
1.917380349666 0.968
1.94134760403682 0.968
1.96531485840765 0.968
1.98928211277847 0.968
2.0132493671493 0.968
2.03721662152012 0.968
2.06118387589095 0.982666666666667
2.08515113026177 0.982666666666667
2.1091183846326 0.989333333333333
2.13308563900342 0.989333333333333
2.15705289337425 0.989333333333333
2.18102014774507 0.989333333333333
2.2049874021159 0.989333333333333
2.22895465648672 0.989333333333333
2.25292191085755 0.989333333333333
2.27688916522837 0.989333333333333
2.3008564195992 0.989333333333333
2.32482367397002 0.989333333333333
2.34879092834085 0.989333333333333
2.37275818271167 0.997333333333333
2.3967254370825 1
};

\addplot [blue, line width=1.2pt,
mark options={solid, opacity=0.4}]
table {%
0 0 
0.023967254370825 0.6
0.04793450874165 0.6
0.071901763112475 0.621333333333333
0.0958690174832999 0.642666666666667
0.119836271854125 0.642666666666667
0.14380352622495 0.642666666666667
0.167770780595775 0.642666666666667
0.1917380349666 0.676
0.215705289337425 0.676
0.23967254370825 0.714666666666667
0.263639798079075 0.761333333333333
0.2876070524499 0.761333333333333
0.311574306820725 0.761333333333333
0.33554156119155 0.761333333333333
0.359508815562375 0.761333333333333
0.3834760699332 0.761333333333333
0.407443324304025 0.761333333333333
0.43141057867485 0.761333333333333
0.455377833045675 0.761333333333333
0.4793450874165 0.761333333333333
0.503312341787325 0.770666666666667
0.52727959615815 0.782666666666667
0.551246850528975 0.782666666666667
0.5752141048998 0.782666666666667
0.599181359270625 0.782666666666667
0.62314861364145 0.782666666666667
0.647115868012274 0.793333333333333
0.6710831223831 0.793333333333333
0.695050376753924 0.793333333333333
0.719017631124749 0.793333333333333
0.742984885495574 0.793333333333333
0.766952139866399 0.798666666666667
0.790919394237224 0.798666666666667
0.814886648608049 0.798666666666667
0.838853902978874 0.892
0.862821157349699 0.892
0.886788411720524 0.905333333333333
0.910755666091349 0.914666666666667
0.934722920462174 0.914666666666667
0.958690174832999 0.914666666666667
0.982657429203824 0.914666666666667
1.00662468357465 0.914666666666667
1.03059193794547 0.921333333333333
1.0545591923163 0.921333333333333
1.07852644668712 0.921333333333333
1.10249370105795 0.956
1.12646095542877 0.956
1.1504282097996 0.956
1.17439546417042 0.956
1.19836271854125 0.956
1.22232997291207 0.956
1.2462972272829 0.956
1.27026448165372 0.956
1.29423173602455 0.956
1.31819899039537 0.956
1.3421662447662 0.973333333333333
1.36613349913702 0.973333333333333
1.39010075350785 0.973333333333333
1.41406800787867 0.973333333333333
1.4380352622495 0.973333333333333
1.46200251662032 0.973333333333333
1.48596977099115 0.973333333333333
1.50993702536197 0.973333333333333
1.5339042797328 0.973333333333333
1.55787153410362 0.989333333333333
1.58183878847445 0.989333333333333
1.60580604284527 0.989333333333333
1.6297732972161 0.989333333333333
1.65374055158692 0.989333333333333
1.67770780595775 0.989333333333333
1.70167506032857 0.989333333333333
1.7256423146994 0.989333333333333
1.74960956907022 0.989333333333333
1.77357682344105 0.989333333333333
1.79754407781187 0.989333333333333
1.8215113321827 0.989333333333333
1.84547858655352 0.989333333333333
1.86944584092435 0.989333333333333
1.89341309529517 0.989333333333333
1.917380349666 0.989333333333333
1.94134760403682 0.989333333333333
1.96531485840765 0.989333333333333
1.98928211277847 0.989333333333333
2.0132493671493 0.989333333333333
2.03721662152012 0.989333333333333
2.06118387589095 0.989333333333333
2.08515113026177 0.989333333333333
2.1091183846326 0.989333333333333
2.13308563900342 0.989333333333333
2.15705289337425 0.989333333333333
2.18102014774507 0.989333333333333
2.2049874021159 0.989333333333333
2.22895465648672 0.989333333333333
2.25292191085755 0.989333333333333
2.27688916522837 0.989333333333333
2.3008564195992 0.989333333333333
2.32482367397002 0.989333333333333
2.34879092834085 0.989333333333333
2.37275818271167 0.997333333333333
2.3967254370825 1
};
\end{axis}

\end{tikzpicture}

%% file: figures/cdf_ul.tikz
\begin{tikzpicture}

\begin{axis}[
width=4in,
height=2.5in,
legend cell align={left},
legend entries={Proposed on, EPA (Proposed off)},
legend style={at={(.52,0.06)}, anchor=south west, draw=white!80.0!black, nodes={scale=1, transform shape}},
tick align=outside,
tick pos=left,
x grid style={white!69!black, dashed},
xlabel={Sum-rate capacity $W$ [bits/c.u.]},
xmajorgrids,
scaled x ticks=false,
xmin=0, xmax=0.04,
xtick={0,0.01,...,0.05},
xticklabels={
  \(\displaystyle {0.00}\),
  \(\displaystyle {0.01}\),
  \(\displaystyle {0.02}\),
  \(\displaystyle {0.03}\),
  \(\displaystyle {0.04}\),
  \(\displaystyle {0.05}\)
},
y grid style={white!69!black, dashed},
ylabel={CDF $F_W$},
ymajorgrids,
ymin=0, ymax=1,
ytick={0,0.2, ..., 1.2}
]
\addplot [red, line width=1.2pt, 
mark options={solid, opacity=0.4}]
table {%
 0.000727859154780506 0.0533333333333334
0.00115309549009854 0.0546666666666667
0.00157833182541658 0.088
0.00200356816073462 0.209333333333333
0.00242880449605265 0.270666666666667
0.00285404083137069 0.270666666666667
0.00327927716668873 0.270666666666667
0.00370451350200676 0.312
0.0041297498373248 0.312
0.00455498617264284 0.350666666666667
0.00498022250796087 0.353333333333333
0.00540545884327891 0.402666666666667
0.00583069517859695 0.452
0.00625593151391498 0.452
0.00668116784923302 0.452
0.00710640418455106 0.457333333333334
0.00753164051986909 0.474666666666667
0.00795687685518713 0.553333333333334
0.00838211319050517 0.605333333333334
0.0088073495258232 0.634666666666667
0.00923258586114124 0.665333333333334
0.00965782219645928 0.665333333333334
0.0100830585317773 0.665333333333334
0.0105082948670954 0.672000000000001
0.0109335312024134 0.672000000000001
0.0113587675377314 0.722666666666667
0.0117840038730495 0.757333333333334
0.0122092402083675 0.757333333333334
0.0126344765436855 0.757333333333334
0.0130597128790036 0.772000000000001
0.0134849492143216 0.781333333333334
0.0139101855496396 0.781333333333334
0.0143354218849577 0.781333333333334
0.0147606582202757 0.801333333333334
0.0151858945555938 0.801333333333334
0.0156111308909118 0.818666666666667
0.0160363672262298 0.828000000000001
0.0164616035615479 0.828000000000001
0.0168868398968659 0.828000000000001
0.0173120762321839 0.828000000000001
0.017737312567502 0.828000000000001
0.01816254890282 0.828000000000001
0.018587785238138 0.860000000000001
0.0190130215734561 0.860000000000001
0.0194382579087741 0.861333333333334
0.0198634942440922 0.877333333333334
0.0202887305794102 0.930666666666667
0.0207139669147282 0.930666666666667
0.0211392032500463 0.930666666666667
0.0215644395853643 0.930666666666667
0.0219896759206823 0.930666666666667
0.0224149122560004 0.933333333333334
0.0228401485913184 0.936
0.0232653849266365 0.936
0.0236906212619545 0.936
0.0241158575972725 0.936
0.0245410939325906 0.952
0.0249663302679086 0.952
0.0253915666032266 0.952
0.0258168029385447 0.952
0.0262420392738627 0.966666666666667
0.0266672756091807 0.966666666666667
0.0270925119444988 0.966666666666667
0.0275177482798168 0.966666666666667
0.0279429846151349 0.966666666666667
0.0283682209504529 0.966666666666667
0.0287934572857709 0.966666666666667
0.029218693621089 0.966666666666667
0.029643929956407 0.966666666666667
0.030069166291725 0.966666666666667
0.0304944026270431 0.966666666666667
0.0309196389623611 0.966666666666667
0.0313448752976791 0.966666666666667
0.0317701116329972 0.982666666666667
0.0321953479683152 0.982666666666667
0.0326205843036333 0.982666666666667
0.0330458206389513 0.982666666666667
0.0334710569742693 0.989333333333334
0.0338962933095874 0.989333333333334
0.0343215296449054 0.989333333333334
0.0347467659802234 0.989333333333334
0.0351720023155415 0.989333333333334
0.0355972386508595 0.989333333333334
0.0360224749861776 0.989333333333334
0.0364477113214956 0.989333333333334
0.0368729476568136 0.989333333333334
0.0372981839921317 0.989333333333334
0.0377234203274497 0.989333333333334
0.0381486566627677 0.989333333333334
0.0385738929980858 0.989333333333334
0.0389991293334038 0.989333333333334
0.0394243656687218 0.989333333333334
0.0398496020040399 0.989333333333334
0.0402748383393579 0.989333333333334
0.040700074674676 0.989333333333334
0.041125311009994 0.989333333333334
0.041550547345312 0.989333333333334
0.0419757836806301 0.997333333333333
0.0424010200159481 0.997333333333333
0.0428262563512661 1
};

\addplot [blue, line width=1.2pt,
mark options={solid, opacity=0.4}]
table {%
0 0 
 0.000428263785403788 0.605333333333334
0.000856526336574115 0.642666666666667
0.00128478888774444 0.642666666666667
0.00171305143891477 0.676
0.0021413139900851 0.761333333333334
0.00256957654125542 0.761333333333334
0.00299783909242575 0.761333333333334
0.00342610164359608 0.761333333333334
0.0038543641947664 0.761333333333334
0.00428262674593673 0.770666666666667
0.00471088929710706 0.782666666666667
0.00513915184827738 0.782666666666667
0.00556741439944771 0.782666666666667
0.00599567695061804 0.793333333333334
0.00642393950178836 0.793333333333334
0.00685220205295869 0.798666666666667
0.00728046460412902 0.798666666666667
0.00770872715529934 0.798666666666667
0.00813698970646967 0.892
0.00856525225764 0.892
0.00899351480881033 0.914666666666667
0.00942177735998065 0.914666666666667
0.00985003991115098 0.914666666666667
0.0102783024623213 0.914666666666667
0.0107065650134916 0.921333333333333
0.011134827564662 0.921333333333333
0.0115630901158323 0.956
0.0119913526670026 0.956
0.0124196152181729 0.956
0.0128478777693433 0.956
0.0132761403205136 0.956
0.0137044028716839 0.956
0.0141326654228542 0.956
0.0145609279740246 0.956
0.0149891905251949 0.956
0.0154174530763652 0.969333333333334
0.0158457156275356 0.973333333333334
0.0162739781787059 0.973333333333334
0.0167022407298762 0.973333333333334
0.0171305032810465 0.973333333333334
0.0175587658322169 0.973333333333334
0.0179870283833872 0.973333333333334
0.0184152909345575 0.973333333333334
0.0188435534857278 0.973333333333334
0.0192718160368982 0.973333333333334
0.0197000785880685 0.989333333333334
0.0201283411392388 0.989333333333334
0.0205566036904091 0.989333333333334
0.0209848662415795 0.989333333333334
0.0214131287927498 0.989333333333334
0.0218413913439201 0.989333333333334
0.0222696538950905 0.989333333333334
0.0226979164462608 0.989333333333334
0.0231261789974311 0.989333333333334
0.0235544415486014 0.989333333333334
0.0239827040997718 0.989333333333334
0.0244109666509421 0.989333333333334
0.0248392292021124 0.989333333333334
0.0252674917532827 0.989333333333334
0.0256957543044531 0.989333333333334
0.0261240168556234 0.989333333333334
0.0265522794067937 0.989333333333334
0.0269805419579641 0.989333333333334
0.0274088045091344 0.989333333333334
0.0278370670603047 0.989333333333334
0.028265329611475 0.989333333333334
0.0286935921626454 0.989333333333334
0.0291218547138157 0.989333333333334
0.029550117264986 0.989333333333334
0.0299783798161563 0.989333333333334
0.0304066423673267 0.989333333333334
0.030834904918497 0.989333333333334
0.0312631674696673 0.989333333333334
0.0316914300208376 0.989333333333334
0.032119692572008 0.989333333333334
0.0325479551231783 0.989333333333334
0.0329762176743486 0.989333333333334
0.033404480225519 0.989333333333334
0.0338327427766893 0.989333333333334
0.0342610053278596 0.989333333333334
0.0346892678790299 0.989333333333334
0.0351175304302003 0.989333333333334
0.0355457929813706 0.989333333333334
0.0359740555325409 0.989333333333334
0.0364023180837112 0.989333333333334
0.0368305806348816 0.989333333333334
0.0372588431860519 0.989333333333334
0.0376871057372222 0.989333333333334
0.0381153682883925 0.989333333333334
0.0385436308395629 0.989333333333334
0.0389718933907332 0.989333333333334
0.0394001559419035 0.989333333333334
0.0398284184930739 0.989333333333334
0.0402566810442442 0.989333333333334
0.0406849435954145 0.989333333333334
0.0411132061465848 0.989333333333334
0.0415414686977552 0.989333333333334
0.0419697312489255 0.997333333333334
0.0423979938000958 0.997333333333334
0.0428262563512661 1
};
\end{axis}

\end{tikzpicture}